\documentclass[sigconf,nonacm]{acmart}
\usepackage{subcaption}
\usepackage{todonotes}
\usepackage{balance}

\begin{document}

\title[The CASTLE 2024 Dataset: Advancing the Art of Multimodal Understanding]{The CASTLE 2024 Dataset:\\Advancing the Art of Multimodal Understanding}

\author{Luca Rossetto}
\orcid{0000-0002-5389-9465}
\affiliation{%
\institution{Dublin City University}
\city{Dublin}
\country{Ireland}
}

\author{Werner Bailer}
\orcid{0000-0003-2442-4900}
\affiliation{%
\institution{JOANNEUM RESEARCH}
\city{Graz}
\country{Austria}
}

\author{Duc-Tien Dang-Nguyen}
\orcid{0000-0002-2761-2213}
\affiliation{%
\institution{University of Bergen}
\city{Bergen}
\country{Norway}
}

\author{Graham Healy}
\orcid{0000-0001-6429-6339}
\affiliation{%
\institution{Dublin City University}
\city{Dublin}
\country{Ireland}
}

\author{Björn Þór Jónsson}
\orcid{0000-0003-0889-3491}
\affiliation{%
\institution{Reykjavik University}
\city{Reykjav\'i{}k}
\country{Iceland}
}

\author{Onanong Kongmeesub}
\orcid{0009-0008-8966-0684}
\affiliation{%
\institution{Dublin City University}
\city{Dublin}
\country{Ireland}
}

\author{Hoang-Bao Le}
\orcid{0009-0000-2496-4347}
\affiliation{%
\institution{Dublin City University}
\city{Dublin}
\country{Ireland}
}

\author{Stevan Rudinac}
\orcid{0000-0003-1904-8736}
\affiliation{%
\institution{University of Amsterdam}
\city{Amsterdam}
\country{Netherlands}
}

\author{Klaus Schöffmann}
\orcid{0000-0002-9218-1704}
\affiliation{%
\institution{Klagenfurt University}
\city{Klagenfurt}
\state{}
\country{Austria}
}

\author{Florian Spiess}
\orcid{0000-0002-3396-1516}
\affiliation{%
\institution{University of Basel}
\city{Basel}
\country{Switzerland}
}

\author{Allie Tran}
\orcid{0000-0002-9597-1832}
\affiliation{%
\institution{Dublin City University}
\city{Dublin}
\country{Ireland}
}

\author{Minh-Triet Tran}
\orcid{0000-0003-3046-3041}
\affiliation{%
\institution{VNU Ho Chi Minh University of Science}
\city{Ho Chi Minh City}
\state{}
\country{Vietnam}
}

\author{Quang-Linh Tran}
\orcid{0000-0002-5409-0916}
\affiliation{%
\institution{Dublin City University}
\city{Dublin}
\country{Ireland}
}

\author{Cathal Gurrin}
\orcid{0000-0003-2903-3968}
\affiliation{%
\institution{Dublin City University}
\city{Dublin}
\country{Ireland}
}

\renewcommand{\shortauthors}{Luca Rossetto et al.}

\begin{abstract}

Egocentric video has seen increased interest in recent years, as it is used in a range of areas.
However, most existing datasets are limited to a single perspective.
In this paper, we present the CASTLE 2024 dataset, a multimodal collection containing ego- and exo-centric (i.e., first- and third-person perspective) video and audio from 15 time-aligned sources, as well as other sensor streams and auxiliary data.
The dataset was recorded by volunteer participants over four days in a fixed location and includes the point of view of 10 participants, with an additional 5 fixed cameras providing an exocentric perspective.
The entire dataset contains over 600 hours of UHD video recorded at 50 frames per second.
In contrast to other datasets, CASTLE 2024 does not contain any partial censoring, such as blurred faces or distorted audio.
The dataset is available via \url{https://castle-dataset.github.io/}.

\end{abstract}



\maketitle

\section{Introduction}

Human interactions and everyday experiences are inherently complex, dynamic, and multifaceted.
Understanding and analysing these interactions is critical for advancing research across numerous fields, including human-computer interaction, social dynamics, psychology, and linguistics.
Although plenty of visual datasets capturing human activities have been created, many of them exhibit significant limitations.
Third-person datasets, for example, often lack the subjective context crucial for interpreting human behaviour,
while first-person datasets frequently limit either the recording duration or scope of activities.
Multi-perspective datasets that combine first-person and third-person views are rare and typically include only a limited number of activities and do not last long enough to capture the full range of interactions and social dynamics characteristic of everyday life.

\begin{figure*}[t]
    \centering
    \includegraphics[width=.9\linewidth]{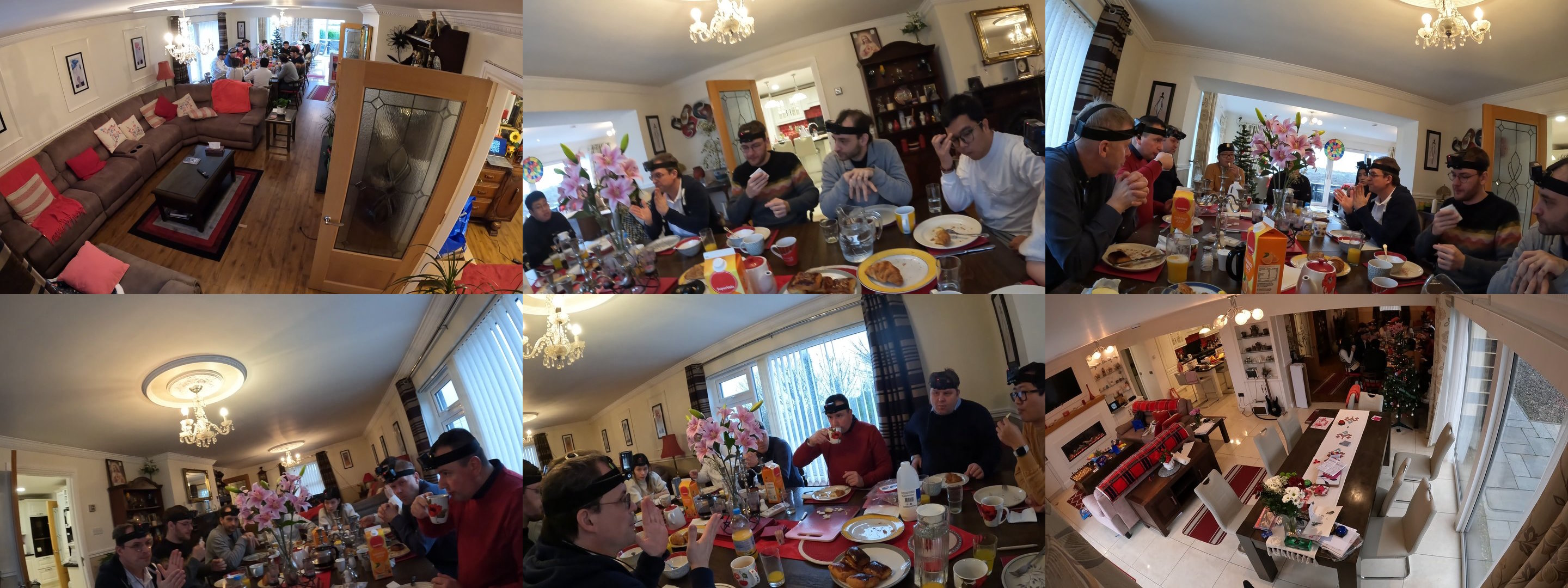}
    \caption{Example from the dataset: six perspectives of a joint breakfast}
    \label{fig:breakfast}
\end{figure*}

\begin{figure*}[t]
    \centering
    \includegraphics[width=.9\linewidth]{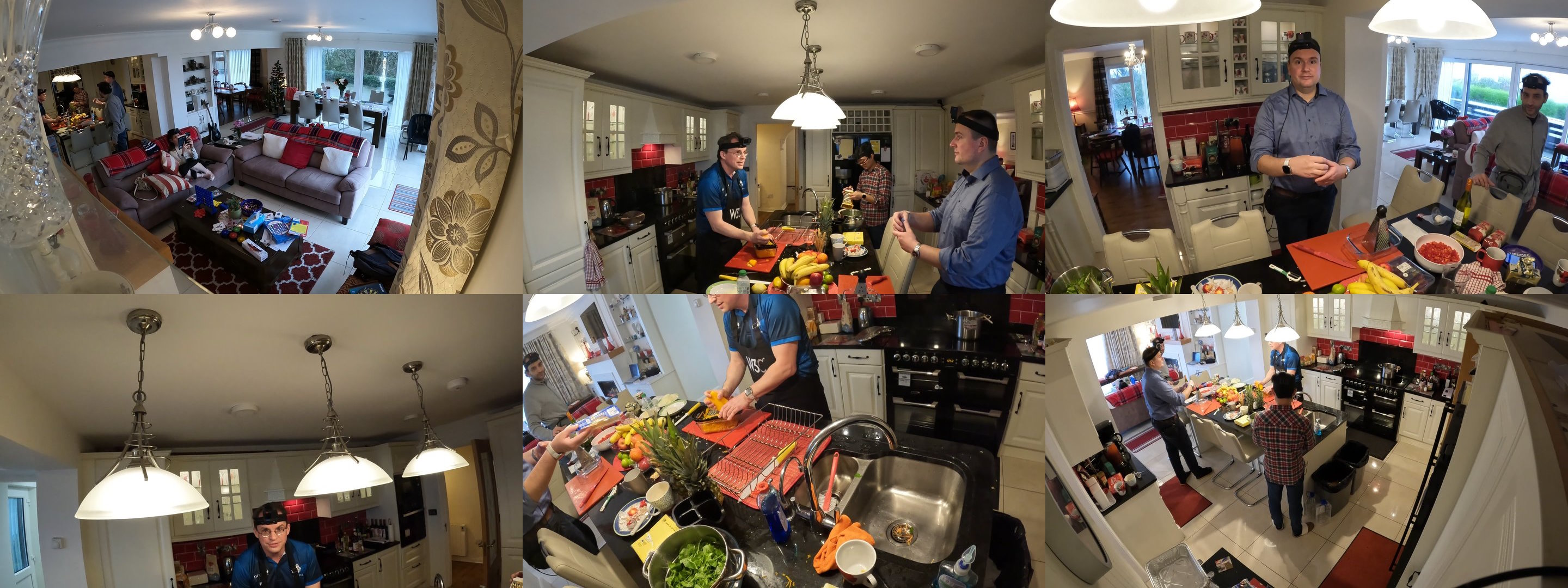}
    \caption{Example from the dataset: six perspectives of cooking activities}
    \label{fig:cooking}
\end{figure*}

\begin{figure*}[t]
    \centering
    \includegraphics[width=.9\linewidth]{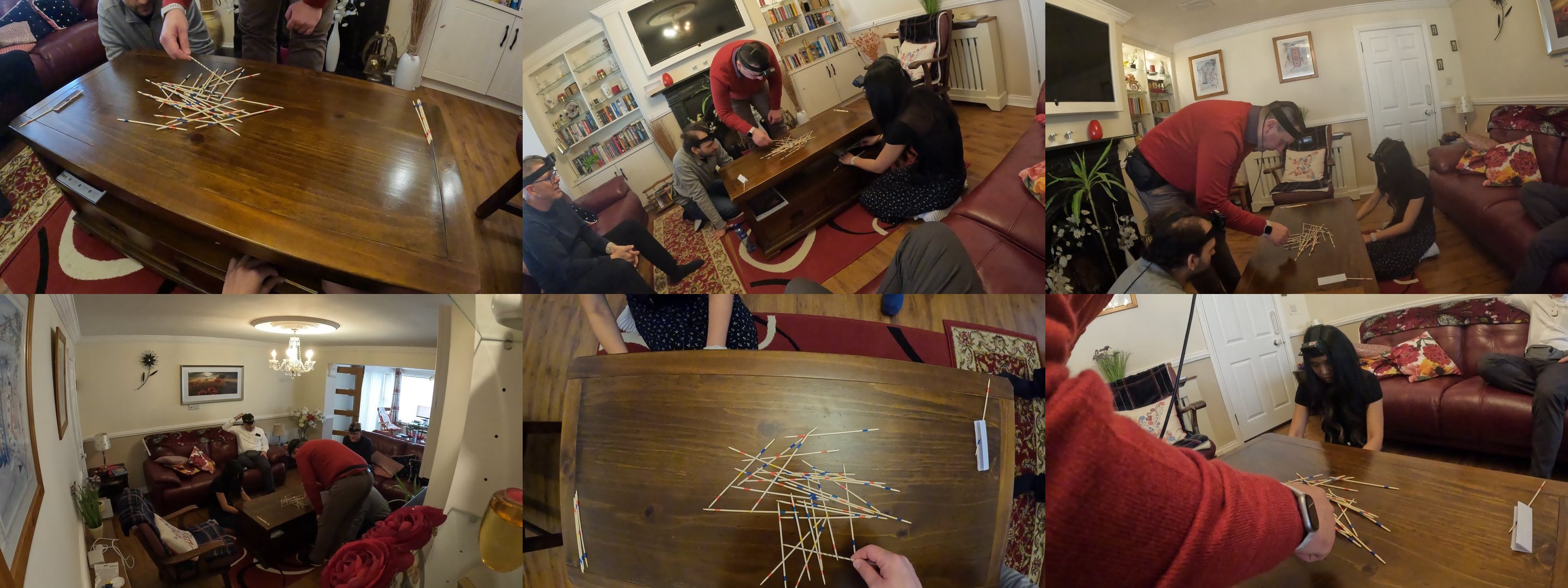}
    \caption{Example from the dataset: six perspectives of people playing a game}
    \label{fig:mikado}
\end{figure*}

In this paper, we introduce the CASTLE 2024 dataset, a multimodal multi-perspective collection of ego-centric (first-person) and exo-centric (third-person) high-resolution video recordings, augmented with additional sensor streams, designed to capture the complexity of daily human experiences.
The dataset captures the experience and daily interaction of ten volunteer participants over the course of four days. It shows a broad range of domestic and social activities, including \textit{cooking}, \textit{eating}, \textit{cleaning}, \textit{meeting} and \textit{leisure activities}, capturing authentic interactions among participants.
The main part of the dataset consists of time-aligned videos from 15 GoPro HERO10 cameras in UHD resolution ($3840\times2160$ pixels) at 50 frames per second, capturing a total of over 600 hours of video and audio data. Ten cameras were worn by the participants, providing immersive ego-centric views, while five stationary cameras offered broader contextual coverage with exo-centric perspectives. 
Figures~\ref{fig:breakfast}, \ref{fig:cooking}, and \ref{fig:mikado} show examples of different situations from multiple perspectives.

Additional metadata recorded by the cameras, such as inertial measurements (IMU) and GPS data is also included in the dataset, providing valuable context for detailed analysis.
In addition, all participants were wearing heart rate monitors and took additional images and videos with additional recording devices, all of which are included as auxiliary data in the dataset.
Moreover, the dataset is also uniquely self-documenting through the inclusion of workshop sessions, during which participants discussed the data generation process and planned further downstream applications. These workshop sessions were carried out in English and touched on a variety of related topics. Although most of the conversations recorded in the dataset are in English, the participants come from diverse ethnic and linguistic backgrounds, and the dataset also contains further conversations in the native languages of some of the participants, including German, Swiss German, and Vietnamese. Such multimodal and multilingual richness makes the dataset particularly interesting for studies in linguistics, social dynamics, and human-computer interaction.

The dataset is expected to be useful for a variety of multimedia analysis and understanding tasks and beyond, such as lifelog retrieval, object and action recognition (including understanding of long, complex and multi-person action sequences), social interaction analysis, and reconstruction and analysis of dynamic 3D scenes. Some of the auxiliary data can already provide metadata for these tasks, while others will require the creation of specific annotations.
Furthermore, the combination of ego-centric and exo-centric perspectives can be used to study the interplay between individual and group activities, as well as the impact of different perspectives on the understanding of human activities and interactions.
Given the rich and diverse nature of the dataset, it is also expected to be useful for the development and evaluation of new methods for multimodal data analysis, including methods for data fusion, multimodal retrieval, and multimodal summarization. By providing a large-scale dataset with a variety of perspectives and sensor streams, we aim to enable researchers to develop and evaluate new methods for understanding human activities and interactions in everyday environments, as well as to explore new applications and use cases for multimodal data analysis.

The remainder of this paper is structured as follows:
Section~\ref{sec:related} discusses other datasets and their commonalities and differences to CASTLE 2024.
Section~\ref{sec:collection} then presents an overview of the methods by which our dataset was constructed.
Section~\ref{sec:properties} details the most important properties of the dataset.
Section~\ref{sec:application} discusses some possible downstream tasks that can be addressed with the dataset.
Finally, Section~\ref{sec:conclusion} concludes the paper.

The dataset is available via \url{https://castle-dataset.github.io/} under a Creative Commons Attribution-NonCommercial-ShareAlike 4.0 International License.

\section{Related Work}
\label{sec:related}
Visual data captured from a first-person perspective has received increased attention in recent years, and several collections have emerged that contain such data in various settings.
An early example is the \emph{LSC Dataset}~\cite{DBLP:conf/mmm/GurrinSJMAHZD19}, used for the annual Lifelog Search Challenge~\cite{DBLP:journals/access/TranNDHSLPNKDJRYAATHSG23}.
It contains a lifelog in the form of an egocentric image sequence representing the first-person view of a single person, covering multiple years of their life.
The images were taken with a cadence of 2-3 images per minute, resulting in a sequence that can be considered a very low frame rate video.

Among egocentric video datasets, a prominent example is \emph{EPIC Kitchens}~\cite{Damen2022RESCALING}, which contains 100 hours of video recorded using head mounted GoPro cameras.
The dataset consists of unscripted activities recorded in 45 different kitchen environments.
Similarly, \emph{HD-EPIC}~\cite{perrett2025hdepic} contains 41 hours of egocentric video recorded in 9 kitchens. It augments the video with eyegaze data and provides full 3D reconstructions of the 9 kitchen environments.

\emph{Ego4D}~\cite{DBLP:conf/cvpr/GraumanWBCFGH0L22} substantially broadened the scope of egocentric video by collecting activity of daily living video with a combined duration of about 3670 hours.
The dataset contains videos of different durations, captured by 931 different camera wearers at 74 locations in 9~different countries.
\emph{MultiEgoView}~\cite{DBLP:conf/nips/HollidtSJHQL024} extends the egocentric view beyond one camera by including recordings from six synchronized cameras worn at different locations on the body of the same person while performing different actions.

Other approaches for understanding scenes and human activities within them use multi-perspective third-person views.
Examples of such datasets include \emph{MM-Office}~\cite{DBLP:conf/icassp/YasudaOSH22} which places four cameras and eight microphones at different locations in an office environment.
\emph{MEVA}~\cite{DBLP:conf/wacv/CoronaOCH21} expands on this theme by compiling $9\,300$ hours of security camera footage from 38 RGB and thermal IR cameras from an access-controlled campus.
The dataset features roughly 100 participants involved in predefined scenarios as well as the spontaneous background activities, recorded over three weeks.

A combination of egocentric and exocentric videos can be found in other datasets such as \emph{Assembly101}~\cite{DBLP:conf/cvpr/SenerCSHSWY22}, which contains multiple time-aligned videos of a single person performing specific assembly and disassembly tasks.
The dataset consists of 513 hours of video recorded by four head-mounted cameras and eight fixed external cameras in a highly constrained environment.
\emph{Ego-Exo4D}~\cite{DBLP:conf/cvpr/GraumanWTKMAABB24} brings the combination of first- and third-person perspective video to a much broader range, providing 1286 total hours of video from a range of different activities in 123 different natural scene contexts.
Individual videos are between 1 and 42 minutes long and are augmented with a range of additional sensor data, including multichannel audio, eye gaze, 3D point clouds, camera poses, and IMU data.

To go beyond single-person actions, multi-perspective multi-user recordings are necessary to capture the complexity of human interactions.
Examples of data collections addressing this include \emph{HoloAssist}~\cite{DBLP:conf/iccv/WangKRPCABFTFJP23}, a large dataset including the recording of interactions between pairs of participants, collaboratively solving a set of predefined object-centric manipulation tasks.   
Every activity is jointly performed by two people with two distinct roles: the `performer' wears a head-mounted recording device and performs a specific activity, while the `instructor' watches and provides instructive information.
The \emph{Aria Everyday Activities Dataset}~\cite{DBLP:journals/corr/abs-2402-13349} also contains the perspective from multiple participants, but does not assign specific roles to them.
It consists of 7.3 hours of video recorded by wearable cameras in 5 locations, with one to two wearers per location.
The videos are augmented with eyegaze and IMU data.

While all of these datasets provide insights into human activities in multiple ways and some of them are augmented with multiple additional sensor streams, none of them provides a long-form multi-perspective representation capturing the richness of everyday activity.
The one dataset that is closest to ours in terms of addressing these challenges is the very recent \emph{EgoLife}~\cite{yang2025egolife}, which was created in a comparable setting to our dataset.
It consists of recordings by 6 participants wearing recording glasses who lived together in the same location for 7 days.
Their location was also equipped with 15 stationary cameras, which provided first- and third-person perspectives.
The egocentric videos are recorded in a square $1408\times1408$ pixel resolution at 20 frames per second and include audio, capturing conversations between the participants.
The videos are augmented with additional eyegaze and IMU data, and the dataset contains additional information, such as a complete 3D reconstruction of the environment, as well as extensive annotations.

Compared to CASTLE 2024, EgoLife covers a longer time span (i.e. 7 vs. 4 days) and includes video from more stationary cameras (i.e. 15 vs 5).
It also contains more explicit annotations of activities which have no direct equivalent in the initial release of CASTLE 2024.
Our dataset does contain more egocentric perspectives (10 vs 6) and recorded video at a higher resolution and frame rate (UHD@50fps vs $1408\times1408$ pixel @ 20fps).
{CASTLE 2024} also contains longer continuously recorded video and more video overall.
In contrast to EgoLife, CASTLE 2024 does not blur faces nor partially anonymise any other content.

\section{Data Collection}
\label{sec:collection}

In the following, we discuss the data collection procedure.

\subsection{Recording Setup}
\begin{figure}
    \centering
    \includegraphics[width=\linewidth]{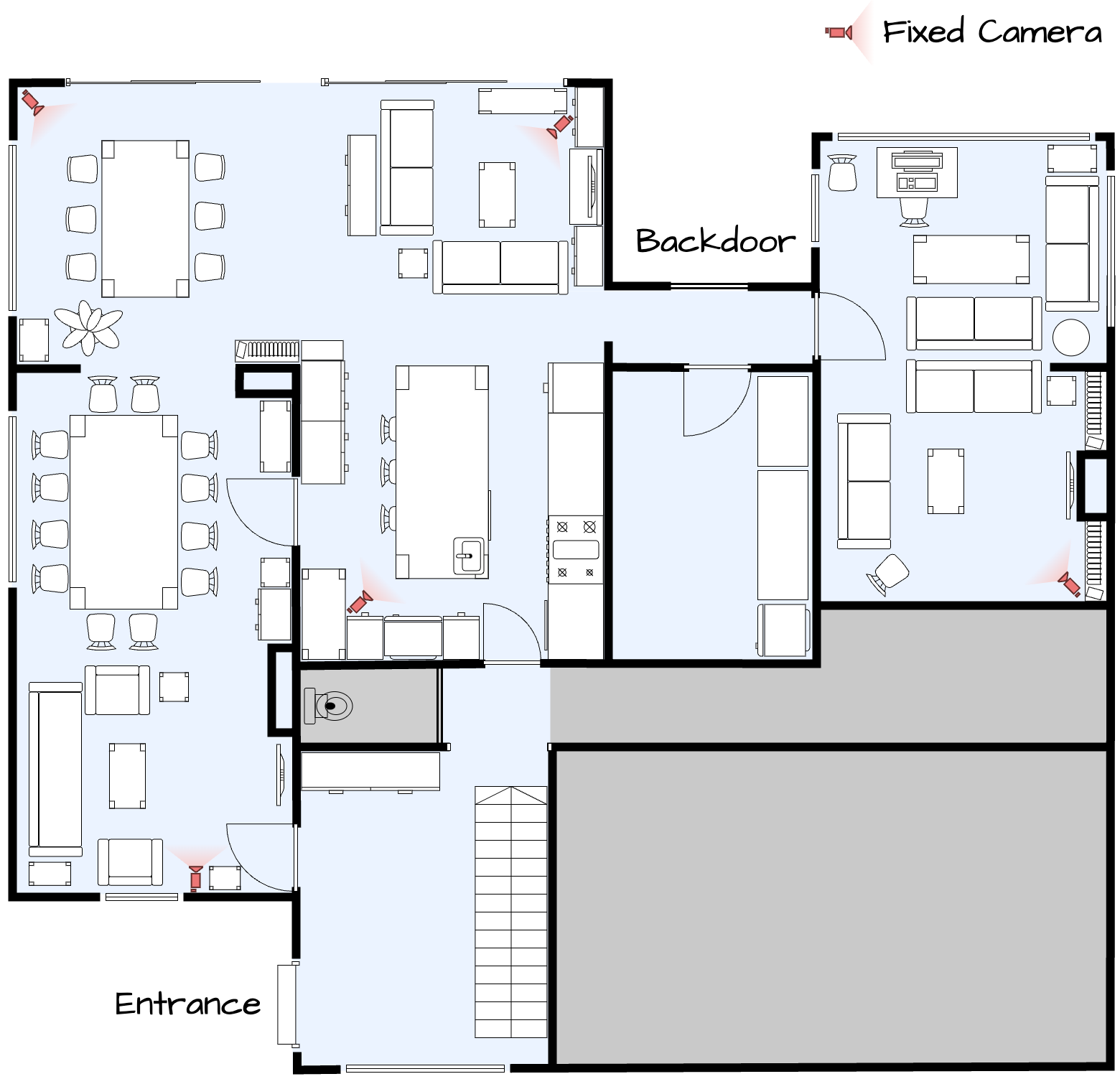}
    \caption{
        Schematic floor plan of the House (not to scale).
        Locations of static cameras are highlighted in red.
    }
    \label{fig:floorplan}
\end{figure}

The CASTLE 2024 dataset was recorded in early December 2024 in a vacation home at the west coast of Ireland, located in a remote area chosen to minimize interaction with third parties.
The recording process included 12 volunteer participants, i.e., the authors of this paper\footnote{Two authors could not be physically present due to scheduling conflicts.} who had all previously agreed to share the data recorded during this time period.
Most of the activities shown in the dataset took place in the common area of the house, schematically depicted in Figure~\ref{fig:floorplan}, which includes a kitchen, a dining area, and various spaces for leisure activities.
Individual bedrooms in the house, as well as a secondary house with additional bedrooms and a common space, served as camera-free zones for privacy.
Some activities were also recorded outside the house, such as walks and visits to a nearby beach, including driving to and from the location. However, the majority of the data was recorded indoors due to the weather conditions at the time of the recording.
The on-site team included ten active \emph{data gatherers} and two \emph{helpers}.\footnote{One helper acted as a data gatherer for the last day of the recording period, since another participant had to leave a day early.}

Each data gatherer was wearing a head-mounted GoPro HERO10 Black, equipped with a 256GB SD-card and connected to an external 20,000mAh battery bank.
The cameras were configured to record in UHD ($3840\times2160$ pixels) resolution at 50 frames per second with minimal motion stabilization.
Each data gatherer was also wearing a FitBit that continuously recorded the wearer's heart rate.

Five stationary cameras of the same make and model as used by the data gatherers were placed at fixed locations in the house, as illustrated in Figure~\ref{fig:floorplan}.
These cameras were powered by an external mains-power adapter and equipped with a 512GB SD-card\footnote{Except for the camera in the reading room, which had a 256GB SD-card.} to support continuous recording throughout the entire day. The cameras were configured to record in the same resolution and frame rate as the data gatherers' cameras.

At the end of every day, SD cards and battery banks were collected for data copying and recharging.
The FitBit devices were worn continuously by the data gatherers.

\subsection{Activities}
The participants were encouraged to participate in a variety of activities. A list of suggested activities was provided to  participants, and was discussed, modified, and expanded on during the recording period.
However, participants were free to participate in any activity they chose. Examples of such activities include cooking, eating, cleaning, reading, playing music instruments, painting and drawing, playing board games, watching television, and most importantly interacting with other participants. 

Some social activities were planned in advance to ensure that multiple participants were present and interacting at the same time.
For example, each day, the participants gathered for a workshop session to discuss the data collection process, plan further activities, and address any issues that arose during the recording. These sessions were recorded and are included in the dataset, acting as a form of self-documentation.
Other social activities that were recorded include reading experiments conducted by one participant for a research project, where eyetracking data was collected using a Gazepoint\footnote{\url{https://www.gazept.com}} GP3 HD eyetracker.
`Happy Quiz', as called by the participants, was also played most evenings, with a quizmaster asking questions to the other participants, who would press a buzzer to answer.

\subsection{Privacy and Consent}
One of the most important aspects of the dataset was that we did not wish to anonymise the participants, for example by blurring faces or distorting voices. Instead, we aimed to capture authentic experiences and interactions among the participants, including their facial expressions, body language, and tone of voice. This decision was made to preserve the richness and authenticity of the data, as well as to enable studies on social interactions, language use, and various other aspects of human behaviour, especially in the context of multimodal data analysis and understanding.
However, we also wanted to ensure that the participants' privacy was protected and that they felt comfortable and safe during the recording process.

Prior to their agreement to participate in the data collection, the participants were informed about the recording process, the intended use of the data, and the measures taken to protect their privacy---such as excluding bedrooms from the recording area and allowing them to deactivate the cameras at any time. Each participant signed a consent form detailing these aspects.
Additionally, participants were given the opportunity to review the data recorded by their own camera and request the removal of specific segments, including conversations or activities they preferred to keep private, before the data was shared publicly.

Interactions with third parties were kept to a minimum and any third party instances that appeared in the recordings were removed to protect their privacy. When third parties were present, the helpers were responsible for ensuring that the cameras were deactivated to prevent recording as needed. Mostly, these occurred when some participants left the house and ventured onto the public road.

\section{Dataset Properties}
\label{sec:properties}

\begin{figure}[t]
    \centering
    \begin{tabular}{cc}
        \includegraphics[width=.45\linewidth]{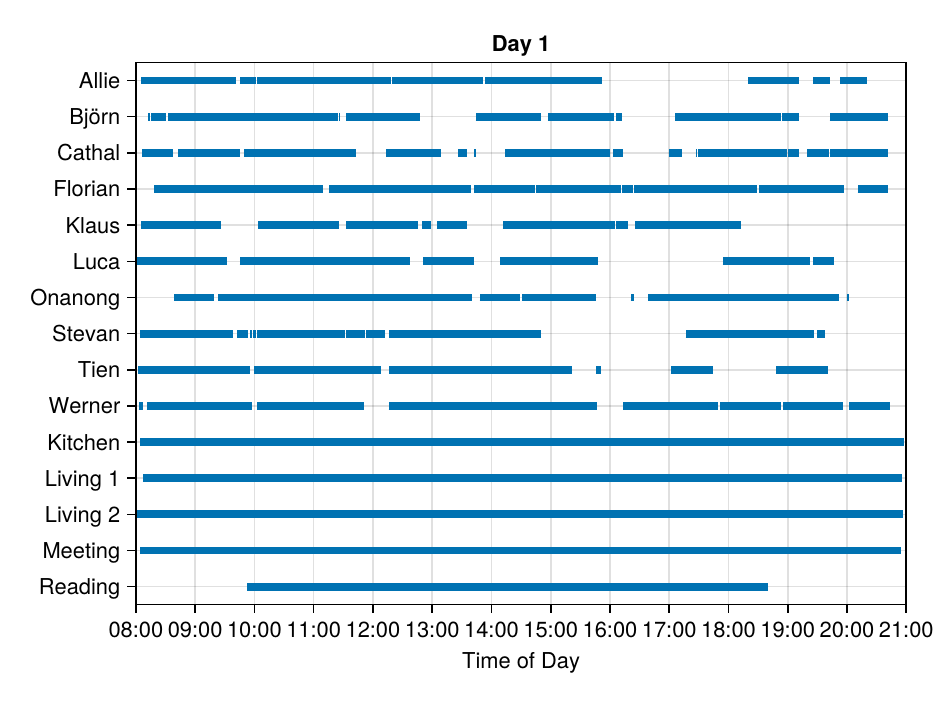}
         &
         \includegraphics[width=.45\linewidth]{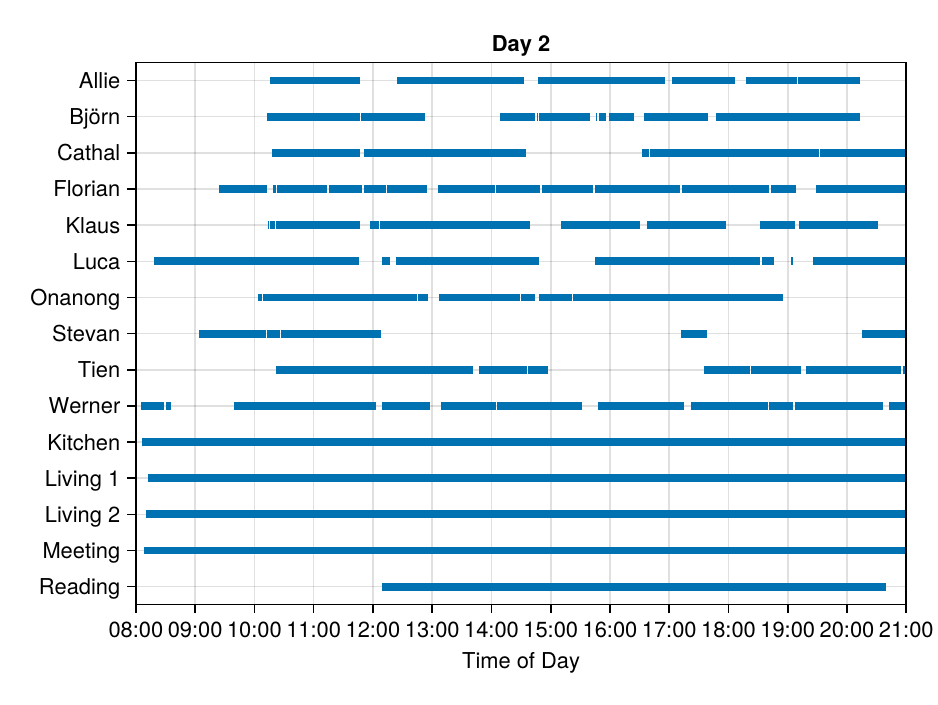}
         \\
         \includegraphics[width=.45\linewidth]{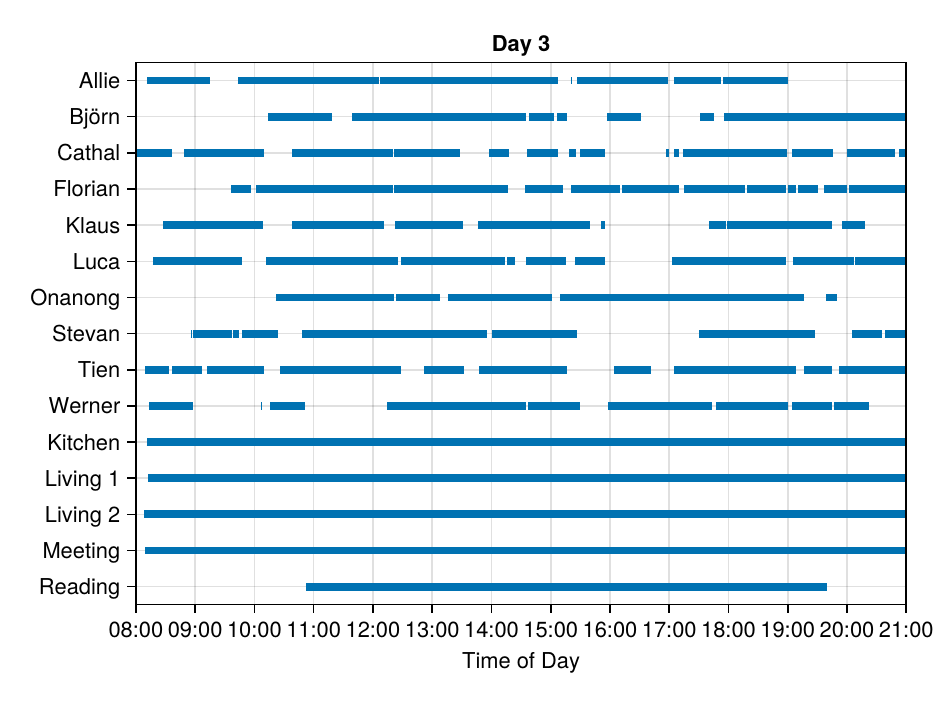}
         &
         \includegraphics[width=.45\linewidth]{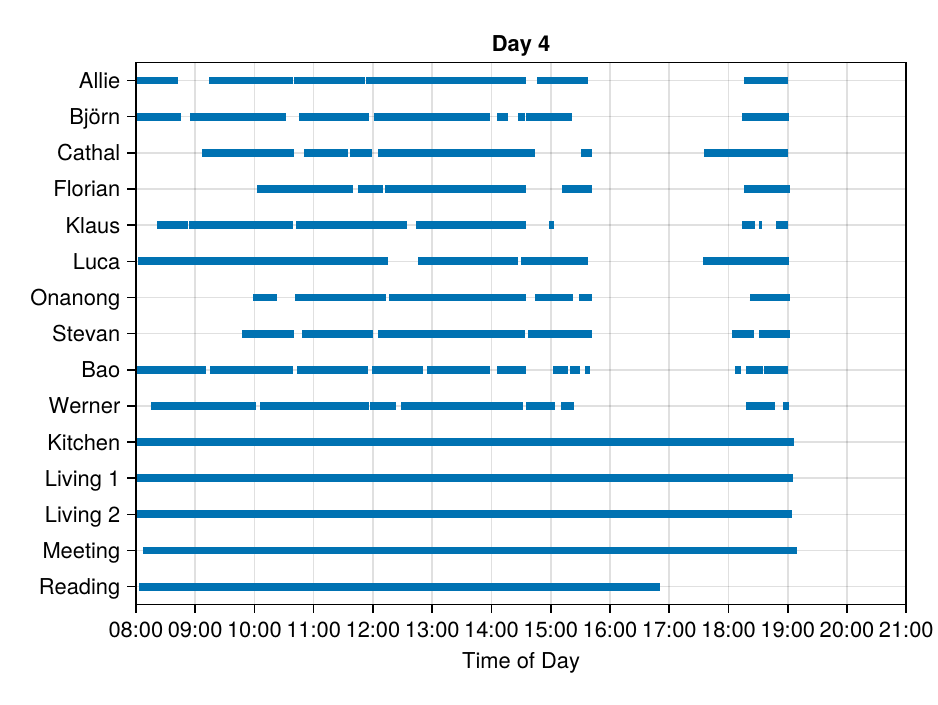}
    \end{tabular}

    \caption{Daily coverage of the individual camera sources of the four days of the dataset}
    \label{fig:recording ranges}
\end{figure}

The CASTLE 2024 dataset consists of two sub-parts -- the main part and the auxiliary part --  totaling 8.22TB in size.

\subsection{Main Part}

\begin{figure}[t]
    \centering
    \includegraphics[width=\linewidth]{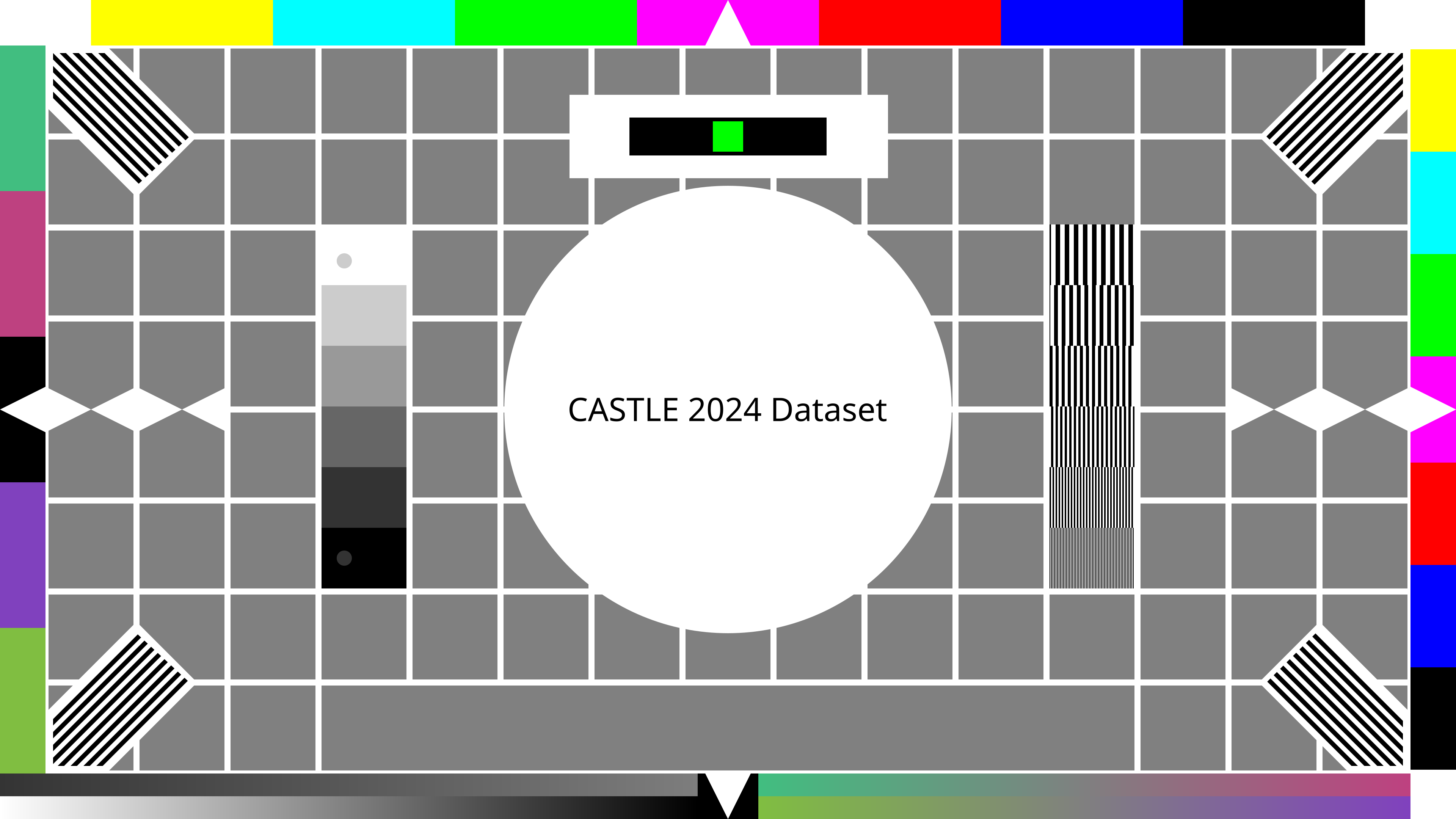}
    \caption{Placeholder used to pad videos when no recording data was available.}
    \label{fig:padding}
\end{figure}

All data in the main part is segmented into one-hour-long, time-aligned segments.
It consists of 666 videos, including their audio streams, and additional files for the other aligned sensor data.
All segments are exactly one hour long (i.e., $180\,000$ frames) and start on the hour, covering the time span from 8am to 9pm, if any recording is available for this time period.
Figure~\ref{fig:recording ranges} shows the ranges during which data was recorded for every stream.
To ensure alignment, gaps in the recording are padded using a placeholder image based on a television broadcast test pattern,\footnote{\url{https://github.com/edent/SVGtestcard}} as illustrated in Figure~\ref{fig:padding}.
Whenever there was no data available for the entire hour, it was omitted completely and replaced with an empty placeholder file to signify the absence of data.
Next to the video files, other sensor streams are made available in CSV format with one file per data stream.
For location data, an additional file in GPS Exchange Format (GPX) is added whenever positioning information was available.
We also provide automatically generated transcriptions of all spoken dialog, generated using a \emph{Whisper V3 Large}~\cite{DBLP:conf/icml/RadfordKXBMS23} model.

\subsection{Auxiliary Part}

The auxiliary part of the dataset contains additional data that is not structured in time-aligned one-hour chunks.
This part contains the following:

\begin{itemize}
    \item Heart-rate of all participants captured once per minute by the FitBit watch.
    \item Images and videos captured by the participants using their personal devices.
    \item Images captured at regular intervals by one participant using a Narrative Clip\footnote{\url{https://getnarrative.com/} PoV wearable camera}
    \item Images captured using a thermal camera, which participants used sporadically.
    \item Eyegaze data captured using a Gazepoint GP3 HD during dedicated reading sessions for some participants.
\end{itemize}

\section{Potential Applications}
\label{sec:application}

We believe that the CASTLE 2024 dataset has the potential to be used in a wide range of downstream applications.
Understanding how researchers might leverage this dataset to address specific tasks provides valuable direction and encourages innovation.
To this end, we outline three initial tasks included in a ACM Multimedia 2025 Grand Challenge:\footnote{\url{https://www.acmmm2025.org}} event instance search, object instance search, and question answering. These represent foundational tasks for advancing multimedia analytics and highlight the dataset's strengths and diversity. Then, we discuss some further potential applications of the dataset.

\subsection{Event Instance Search}
\paragraph{Motivation}
Being able to search for specific events in a large-scale multimodal dataset is a fundamental task in multimedia retrieval.
In the context of the CASTLE 2024 dataset, event search can be used to find specific activities or interactions among the participants.
For example, one might want to find all instances of a specific action, such as someone making coffee, or all instances of a specific interaction, such as someone telling a joke and the others laughing. Being able to search for such events can be the first step towards more complex tasks and applications.

\paragraph{Task Definition}
Given a textual query describing an event in natural language, such as `someone making coffee' or `someone telling a joke and the others laughing', the task is to retrieve all video segments in the dataset that contain the relevant event. The events are to be identified by the time range and video ID\@.
The task can be evaluated using standard information retrieval metrics, such as mean average precision (mAP) or recall at $k$.

\subsection{Object Instance Search}
\paragraph{Motivation}
Object instance search is another fundamental task in multimedia retrieval, where the goal is to find all instances of a specific object in a large-scale multimodal dataset. This specific object can be described by a textual query or an image of the object.
The CASTLE 2024 dataset contains a wide variety of objects, such as kitchen utensils, food items, books, and small decorative items which were moved around by the participants during the recording period.
This task can be challenging due to the large number of objects in the dataset, the diversity of the scenes in which they appear, and the sheer volume of data.

\paragraph{Task Definition}
Given a natural language query text describing an object, such as `a cookie cutter shaped like a star', or a reference image of the object, the task is to retrieve all occurrences of that object in any of the video streams.
Similar to event search, the task can be evaluated using standard information retrieval metrics.

\subsection{Question Answering}
\paragraph{Motivation}
It is human nature to ask questions about the world around us. Recent advances in natural language processing have enabled the development of question answering systems that can answer questions about text, images, and videos.
Video question answering is a challenging task that requires understanding of both the visual and textual content of the video.
Participants in the CASTLE 2024 dataset engaged in a wide range of activities and interactions, providing a rich source of data for video question answering. The long-form nature of the videos and the multiple perspectives captured by the cameras make this task particularly challenging and interesting.

In contrast to how most video question answering tasks are defined, where a video is provided with a question, we aim to elevate the challenge to a more general format, similar to the question answering task in the Lifelog Search Challenge~\cite{DBLP:journals/access/TranNDHSLPNKDJRYAATHSG23}.
This means that the question is unbound to any specific video and can refer to anything that happened during the recording period and was captured in visual or audio channel of at least one camera.
As such, the task is more challenging as the answer must be found by searching through the entire dataset.

\paragraph{Task Definition}
Given a question formulated in natural language, the task is to find an answer to the question. Answers are to be provided in natural language and include references to sensor streams and time intervals to provide evidence.
The task can be evaluated using standard question answering metrics, such as accuracy or F1 score.

\subsection{Other Potential Applications}

Beyond the tasks outlined above, the richness of the CASTLE 2024 dataset can be useful for numerous other research directions.
Its multi-perspective visual coverage is particularly well-suited for training and evaluating computer vision models for tasks such as activity recognition, object detection and multi-perspective object tracking, or scene reconstruction.
Similarly, the synchronised multimodal data streams offer opportunities to develop advanced techniques for multimodal data analysis, including data fusion, cross-modal retrieval, multimedia forensics, and multimodal summarisation.
The detailed capture of social interactions and activities can be a valuable resource for investigating social dynamics, language use, and the development of sophisticated human-computer interaction methodologies. This also applies to AI-based linguistic research, where the dataset's multilingual conversational data can be used to develop and evaluate models for automatic speech recognition, multilingual language modelling, speaker identification, dialect analysis, and sentiment (or humour) detection across languages and conversational contexts.

\section{Conclusion}
\label{sec:conclusion}

In this paper, we presented the CASTLE 2024 dataset, a multimodal, multi-perspective collection comprising synchronized high-resolution ego-centric and exo-centric video recordings, complemented by additional sensor data, aimed at comprehensively capturing daily human experiences.
This data set documents authentic interactions and diverse domestic and social activities -- including cooking, eating, cleaning, and leisure activities -- among ten volunteer participants over four days.

Although the initial release of the dataset does not include in-depth semantic annotations, we foresee that CASTLE 2024 might still prove useful in a range of applications.
In particular, the presented dataset goes beyond currently available alternatives in terms of content diversity and concurrent points of view.
It also increases authenticity and representativeness by forgoing any partial censoring of recordings.

\begin{acks}

This work was partially supported by an ACM SIGMM Special Initiatives Grant.
Individual contributors were further supported by: the Swiss National Science Foundation through project MediaGraph (contract no.\ 202125) and project \href{https://data.snf.ch/grants/grant/193788}{``Participatory Knowledge Practices in Analog and Digital Image Archives''} (contract no.\ 193788); by Research Ireland under grant number 13/RC/2106 P2; and by the Icelandic Research Fund grant 239772-051.

\end{acks}

\bibliographystyle{ACM-Reference-Format}
\balance
\bibliography{bibliography}

\end{document}